\documentstyle[11pt,newpasp,twoside]{article}
\def\edcomment#1{\iffalse\marginpar{\raggedright\sl#1\/}\else\relax\fi}
\reversemarginpar

\begin{document}
\title{Extreme scattering of pulsars}
 \author{Mark A. Walker}
\affil{Special Research Centre for Theoretical Astrophysics,\\
School of Physics, University of Sydney, NSW 2006, Australia}

\begin{abstract}
Extreme Scattering Events are radio-wave lensing events caused
by AU-sized concentrations of ionised gas. Although they were
discovered more than a decade ago we still have no clear picture
of the physical nature of the lenses.
To discriminate between the various models, we need to
amass more information on multiple imaging phenomena. 
Pulsars are perfect targets for such studies: they
offer six- to ten-times the information content of quasar
observations,
and their small angular sizes make them sensitive to distant
lenses. In addition, small source-size means that multiple
imaging can be studied even when there is little change
in the total source flux, because the (weak) secondary images
interfere with the primary and create periodic
fringes in the spectrum. 
\end{abstract}

\section{Introduction}
Extreme Scattering Events (ESEs) were discovered as a result of
long-term monitoring of the radio fluxes of compact
radio quasars (Fiedler et al 1987). The events themselves are
characterised by dramatic,
frequency-dependent flux variations occurring within a period
of a month or two. It is broadly agreed (Fiedler et al 1987, 1994;
Romani, Blandford \& Cordes 1987; Clegg, Chernoff \& Cordes 1988)
that ESEs are a due to ionised gas drifting across the line-of-sight,
and that the transverse dimensions of the ionised region are only a
few AU. If this scale is also representative of the longitudinal
dimension of the ionised gas -- i.e. it takes the form of a
quasi-spherical blob -- then the mere existence of such regions
presents us with a severe problem: the necessary electron
density is $\sim10^3\;{\rm cm^{-3}}$ and at $10^4$~K this implies
pressures $\sim10^3$ times larger than is typical of the
Interstellar Medium (ISM). How can such regions exist?

One possible solution to this conundrum was advanced by
Walker \& Wardle (1998): a neutral gas cloud of radius $\sim$~AU
would develop a photo-ionised wind, as a consequence of the
ambient Galactic radiation field, with about the right
electron density at the ionisation front. Furthermore, the
light-curves for such a lens offer a good representation
of the data for the best known example of an ESE.
However, if this interpretation is
correct, then it follows that the neutral gas clouds must
comprise a large fraction of the mass of the Galaxy, i.e.
they form a major component of the Galactic dark matter.
The success of this picture as a model for ESEs provides
a strong incentive for studying ESEs in detail, and with
renewed vigour. This paper offers some ideas on how this can
best be achieved, arguing that pulsars are by far the best
targets for such studies.

\section{Observables}
In designing experiments to study lenses, it is helpful to
first set-out what quantities we might be able to observe
during a lensing event.
There are, in principle, six measurable quantities of interest
for each image of any small-diameter source: the size, shape and
orientation of the image; the location of the image; and the
delay of the image. If $n$ is the refractive index, and
$N\equiv\int{\rm d}s\,(1-n)$ is its line-of-sight integral, then
these measurables are determined, respectively, by
$\partial_i\partial_jN$ (three independent quantities; $i,j$
denote the transverse coordinates), $\partial_iN$ (two quantities),
and a combination of $N$ and $(\partial_iN)^2$. The image size
(solid angle) is perhaps the most basic measurable as it
determines the received flux; the image magnification is just the
ratio of apparent source sizes with/without the lens present.
Because the lenses
are only milli-arcseconds in size, measurement of image location,
shape and orientation is only possible with VLBI, if at all. Measurement
of the time delay requires a temporal variation in the intrinsic
brightness of the source, and pulsars are eminently suitable.

An important point about the foregoing is that single dish
(or compact interferometer) observations, which cannot hope to
resolve the individual images on the sky, are enormously more
informative if pulsars are targeted rather than quasars. The
information gain is not simply a factor of two, but rather it
is a factor of two for every image present, and it may be
expected that three or five images (always an odd number)
could be present during an
ESE, so a factor of six or ten is possible in principle.
The reason for this gain is that the presence of a delay allows
the images to be resolved in time, whereas without the time
dimension one only knows the total magnification (i.e. summed
over all images). In practice it might not prove possible to
separate {\it all\/} of the images, but the point remains that pulsars
are much more informative than quasars when it comes
to single-dish observations. In turn this means that we can
distinguish between ESE models far more effectively with
pulsars than with quasars.

\section{Why pulsars are so useful}
The considerations just given provide, in themselves, good reason
to prefer pulsars as targets for studying ESEs. There are two
other considerations which are important: pulsars are
extremely compact, and they are steep-spectrum sources.

Their very small angular size has two benefits. First, the
coherence patch of the radiation field is correspondingly large.
This means that when multiple images are present, one of
their manifestations is periodicity in the dynamic spectrum
--- a consequence of interference between two ray paths. This
phenomenon is of great utility: a relative delay between images
of order a microsecond, say, would be straightforward to observe
as a MHz periodicity (even for glitching pulsars), but would be a
tall order to detect in pulse time-of-arrival residuals (even for
a millisecond pulsar; cf. Cognard et al 1993). Another big advantage
of utilising pulsar dynamic spectra is that one can study very
weak (low magnification) secondary images, because the power
resident in periodic interference fringes is the cross-power
between two images, and is proportional to the geometric mean
of their magnifications. This point has
been beautifully illustrated by Rickett, Lyne \& Gupta's (1997)
study of the dynamic spectra of B0834+06, in which secondary
images of magnification $\sim0.1$\% were revealed.

The second benefit of small angular size is that one retains
maximum sensitivity to lenses of small angular diameter; if
a lens were apparently much smaller than the background source,
then it would have little detectable effect. Compact radio
quasars have angular sizes of order a milli-arcsecond, typically;
this is comparable to the size of the lenses (at a distance of
a few kpc), so the Fiedler et al (1994) data set is
unlikely to have provided a complete census, and could
conceivably have missed a large fraction of the lens population!
In the case of pulsars, the intrinsic angular diameter is so small
that the observed size is always limited by interstellar scattering,
i.e. diffraction caused by density inhomogeneities in the ionised
ISM (e.g. Rickett 1990). This is as good as we can hope to achieve.

The fact that pulsars are steep-spectrum sources is also beneficial
in that they are easily studied at long wavelengths, $\lambda$,
where rays are refracted through large angles. Indeed, because
$N\propto\lambda^2$, the cross-section for multiple imaging
increases roughly as $\lambda^4$ at long wavelengths. Interpretation
of this fact requires care, because multiple imaging events are
usually recognised as such only if they split the images by more
than the angular size of the image, i.e. the scattering disk in
the case of pulsars. (In particular this is true of the spectral
periodicities mentioned earlier.) As the scattering disk size
also scales as $\lambda^2$, roughly, the apparent incidence of
multiple imaging may be wavelength-independent. One point
to bear in mind is that refraction through larger angles means
that power is spread over larger physical areas in the plane of
the observer, so the observed flux is smaller. Thus large lensing
cross-sections go hand-in-hand with secondary images which are,
typically, weak.

\section{Finding events}
One of the main barriers to studying ESEs is that they are
rare events: Fiedler et al (1994) estimated that the lenses cover
only $\sim5\times10^{-3}$ of the (extragalactic) sky, requiring
many source-years of monitoring for every event. The burden
of such a monitoring program can be lightened considerably if
it is acceptable to observe only infrequently, rather than the
daily sampling of Fiedler et al (1994), but if one relied upon
flux monitoring, then the signature of an ESE might well
be missed if sampling were infrequent. Here, again, pulsar
dynamic spectra prove useful: if periodicities are present
then we know, without reference to any other data, that multiple
imaging is occurring; this phenomenon (with weak secondary images)
is expected, in essentially all lens models, as a pre- and post-cursor
of  ESEs. (ESEs can be thought of as multiple imaging events in
which the primary image magnification reaches values far from
unity.) The following strategy thus suggests itself.
Monitor pulsar dynamic spectra for the presence of periodic fringes;
the sampling interval need only be a month or so, but high
spectral resolution is important (1~KHz, say), as the image delays
could be large. If no
periodicity is detected in the spectrum of a particular pulsar,
at a given epoch, then that pulsar would not be expected to
undergo an ESE within the next month. On the other hand,
if periodicity is seen, then multiple images are present at
that epoch, and there is a chance that an ESE will take
place within the next month. It is then appropriate to activate
a program of detailed multi-frequency observations, including
daily acquisition of dynamic spectra, and VLBI. This strategy
allows one to monitor a large sample of pulsars, achieving good
sampling on the ESEs, with only a modest commitment of
telescope time. For example, a total
sample of 100 pulsars could be monitored monthly; a small
fraction of these will display
multiple imaging at any one time, so only a handful of
pulsars need be observed on any given day.
Not all pulsars are equally appropriate for this type of program:
nearby pulsars offer little
chance of the signal encountering a lens between source and
observer; strongly scattered pulsars are not useful because
any lens is unlikely to split the images by an angle larger than
the size of the scattering disk.

\section{Signatures}
The acid test of any lens model, if not the physical model
it derives from, lies in fitting the observed image
properties to predictions calculated from the model $N$.
However, there are also some qualitative aspects
of the lensing behaviour which would help to
determine the correct physical picture. In particular it
is important to note that observations have not yet
established the lens symmetry, if any; some symmetry does
appear to be required because the light-curves are,
crudely, time-symmetric. The natural choice
is axisymmetry, arising from an underlying
spherical symmetry; however, this immediately confronts
us with the implication of an exploding lens, as discussed
in \S1.

Two qualitative tests for axisymmetric lenses can be given.
First, if VLBI observations are able to separate the images
from each other, it should be seen that all images lie on
the same line (of azimuth $\phi$). This line should rotate
systematically with time, $t$, such that $\tan\phi\propto t$,
where both $\phi$ and $t$ are measured from the mid-point
of the event. Secondly, for an axisymmetric lens, the geometry
is effectively stationary at $t=0$, so that all
image magnifications and delays should be quadratic in time
around $t=0$. Even simple tests of this kind could advance
our understanding of ESEs considerably.

\end{document}